# Evaluation of An Indoor Localization Engine

Christophe Villien, Anne Frassati
CEA, LETI
17 avenue des Martyrs, France
Christophe.villien@cea.fr

Bruno Flament,
MOVEA-INVENSENSE
22 AVENUE DOYEN LOUIS WEIL

*Abstract*—Pedestrian Indoor localization based on modalities available in modern smartphones have been widely studied in literature and many of the specific challenges have been addressed. However, very few approaches consider the whole problem and proposed solutions are very often evaluated under very limited scenarios. We propose a fusion engine for localization that makes use of various data provided by a smartphone (Inertial sensors, pressure sensors, Wi-Fi, BLE, GNSS, map etc.) to provide a fused localization that is robust under harsh conditions (poor RSS coverage, device position change etc.). Moreover, our solution has been evaluated for hardware integration and tested over a large database including more than 250 experiments representing different scenarios, showing feasibility of lightweight implementation and good results over various conditions.

*Keywords: Indoor pedestrian localisation, Map matching, Particle Filter, Fused location provider*

I. INTRODUCTION

Indoor localization encompasses a wide variety of scenarios and technologies, each of them facing very different challenges. This paper focuses on smartphones based localization for pedestrians, which is one of the most pervasive and studied technology. Indeed, smartphones have several key advantages over other platforms: they are almost everywhere, concentrate dozens of sensors, come with high processing power, connectivity and display to share or retrieve useful information like maps, and generally host the client application that will exploit position data. However, they lack the killer technology that provides high positioning accuracy (e.g. <1m) in any environment. Integration of a localization engine, which we will refer to as Fused Localization Provider (FLP) by analogy with the Android terminology, is also challenging since it should run in an always-on, low-power processor generally called sensor hub (SH) and access high level data like e.g. map or beacons' positions.

Beyond those general aspects, technical challenges are:

- **Continuity of service**: FLP should provide seamless positioning while user is moving between places where some sensors could be unavailable. For instance, GNSS is unavailable indoor, some areas have poor Wi-Fi coverage, map is useless in large open spaces etc.

- **Robustness:** FLP should deliver a reliable position in real time for a wide variety of users, devices and venues.

- **Device position change (DPC)**: smartphones are frequently moving from one position on the body to another, for instance from the pocket (while walking) to the hand (for texting). Hence, system should handle those transitions by tracking the misalignment angle (MA) between the device orientation and the user heading.

- **Map representation**: Map representation should support very different types of venues (mall, offices, underground parking lot etc.) and it also have a significant impact on processing complexity. Some "positive" maps (PM) describe all possible motions or positions which could induce some errors due to quantization of state space, increase computational burden and memory requirements. Whereas "negative" maps (NM) that describes forbidden transitions (e.g. walls) are more compact but less informative as they do not help to find possible next states using privileged directions for instance.

- **Floor changes**: floor changes are commonly detected using fast pressure changes or based on radio signals (e.g. Wi-Fi, BLE). However, beyond floor changes detection itself, localization during transition phase is rarely handled specifically. Indeed, walk in stairways for instance defeats classical Pedestrian Dead Reckoning (PDR) algorithms due to different gait models and is also critical for map matching algorithms as they correspond to large displacements in very small areas. Moreover, typical processing of pressure sensors involves a filtering stage that introduces a delay regarding the floor detection which can shift back the position.

Most of those issues have been widely investigated in literature. Continuity of service if handled in various different ways, some approaches fuse map and inertial measurements only [1]-[5], others integrate also RSSI measurements [8],[10],[9] very few consider GNSS [7]. Map representation is mainly based on PM [2],[3],[6],[10],[11], but could also be based on graphs [8],[9] or grids [4],[5]. DPC is considered in [8] using a dedicated algorithm whereas it is processed at the fusion stage in [7]. Several authors have also developed algorithms that support multi-floor maps [6][8][3][9][11].

However, very few authors have proposed an extensive evaluation of their solution. Results are often given for a limited set of experiments - if not a single experiment - and generally tested under the same conditions (e.g. smartphone in hand, no DPC etc.). Another key aspect that is generally missing is the hardware footprint of the proposed solution (CPU burden, memory requirements) which may be unaffordable for real-time implementation in some cases.

Our contribution is mainly the evaluation of a FLP that merges together disparate solutions to address the various aforementioned problems. This evaluation includes extensive testing (257 experiments) under very different conditions, and hardware footprint characterization.

Paper is organized as follow, section II will describe the environment (Sensor Hub) and the architecture for FLP integration, whereas section III will give details about the FLP design. Section IV is devoted to the evaluation of our solution and a conclusion will be given in section V.

## II. SENSOR HUB DESCRIPTION

### A. Architecture

In this study, we assume that the target device has the architecture depicted in Fig. 1. According to this architecture, which shows only sensors and functions involved in localization, all algorithms are running in the SH, in order to reduce the overall power consumption by offloading some computations from the application processor and enabling this latter to sleep for longer periods while maintaining real time tracking. SH is connected to motion sensors (accelerometer, magnetometer and gyro), pressure sensors, GNSS receiver and also to Received Signal Strength (RSS) measurements from Wi-Fi and BLE. FLP can receive high level data like e.g. map, user size or beacons' positions from the application processor.

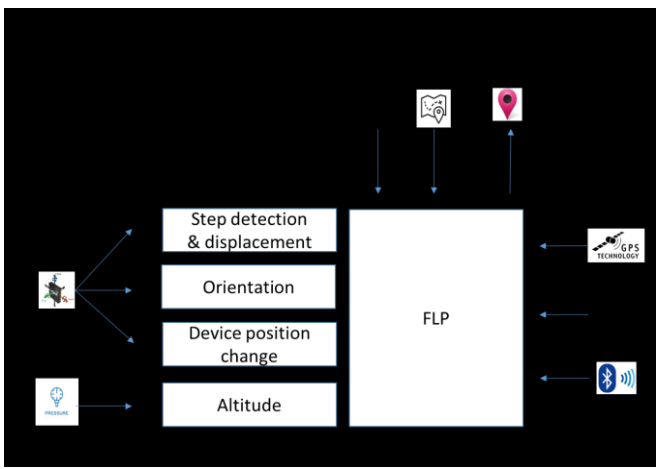

Fig. 1. Typical architecure for FLP implementation

SH also implements some motion processing functions which are part of PDR and which we describe hereafter shortly.

### B. Steps detection and step length

Our steps detection algorithm uses the norm of the accelerometer readings [1] which is invariant to device orientation. This signal is then bandpass filtered in the [1Hz-2Hz] band to extract the fundamental frequency of the walk and remove all spurious harmonics. At this stage the signal is essentially a sinewave and it is quite straightforward to detect steps using max detection. Main limitation of this algorithm are the false alarms (FA) caused by the device manipulation which could be misinterpreted as steps. Thus, steps detection is invalidated when the DPC module indicates that a manipulation is in progress. However, this introduces a new issue when the user is manipulating its smartphone while he is walking. Some heuristics based on the regularity of the walking signal could then be used to decide if the DPC invalidation applies or not.

Steps detection is further converted into displacement using the following step length formula [12]

$$L = a.F + b.H + c \qquad (\text{II-1})$$

Where $L$ is the step length (m), $F$ the step frequency (Hz), $H$ is the user's height (m), $a, b$ and $c$ three constants which we determined using our database and least squares fitting. Values we found are $a=0.339$, $b=0.585$ and $c=-0.923$.

### C. Orientation

Orientation of the device is generally obtained by fusion of accelerometer and gyro (AG) sensors, and eventually includes magnetometer (AGM). Using magnetometer is necessary to compute an absolute orientation with respect to geomagnetic north, however it introduces several difficulties. First, absolute orientation of interest is the one of the user which is related to the one of the device through MA that represents the way the device is carried by the user. Some techniques based on principal component analysis (PCA) [8] of the accelerometer signal can be used for that purpose but those are not always reliable and cost some additional CPU power. Second, magnetometers or their environment (i.e. the device) are subject to magnetization which results in sudden changes of bias and/or scale values. Hence, proper calibration of the sensor should be maintained either through specific procedure on the user side which is limiting, or through autocalibration algorithms but those are not very reliable. Finally, indoors magnetic fields are known [13] to be strongly disturbed. Several algorithms could be used to perform AG or AGM fusion [14], choice being dependent on the tradeoff between performance and CPU cost. A critical point regarding performance is the estimation of the gyro bias, especially the startup bias, which can be very high for MEMs sensors (i.e. typ. 5deg/s) and causes important drifts. This latter could be included within the state vector and estimated along with the orientation, or estimated separately using static phase of the device where the gyro readings correspond directly to the biases.

After comparing several algorithms (e.g. MEKF [14], QUEST [14], Magdwick [15]) we found that best performances were obtained with an Additive Extended Kalman Filter (AEKF) [14], and a separate gyro bias estimator making use of static phases. Although discussions about the orientation algorithm itself are beyond the scope of this paper, we may just say that AEKF is a straightforward implementation of an EKF having a four-dimensional quaternion state vector updated by adding a quaternion error, quaternion being forced to unit norm afterwards. We also choose to compute the orientation of the device based on AG measurements only, to circumvent aforementioned issues of magnetometer. Hence, our orientation module delivers relative orientation, absolute orientation which include the MA being solved at the FLP stage.

*D. Device Position Change*

Architecture of Fig. 1 shows that our FLP makes use of a DPC flag indicating any change of the device position with respect to the user. This information is very useful for the FLP which is responsible for tracking the MA, and also for step detection to remove FA caused by manipulations. Main challenge regarding this function is to discriminate between motions that are inherent to user's trip (i.e. turns) and those that are specific to DPC. Our approach [16] detects variations of the average vertical direction within the body frame. Indeed, this principle is based on the following two basic observations: first, vertical direction remains constant when user makes a turn or goes straight, second, it is unlikely that the vertical directions does not change, at least temporarily, when a smartphone is moved from one position to another on the body (e.g. from pocket to phone call). Vertical direction in the body frame can be obtained easily by averaging accelerometer measurements (e.g. over 1 second) to remove walk components. Any variation between a previously recorded vertical direction $z_{ref}$ and the current direction $z_k$ can be detected by monitoring the angle $\alpha_k$ between the two vectors

$$\alpha_k = sin^{-1}(\|z_{ref} \otimes z_k\|), \qquad (\text{II-2})$$

where $\otimes$ is the cross product. If $\alpha_k$ exceeds a certain threshold (e.g. 20deg) then a DPC is flagged up until the vertical direction stabilize and a new vertical direction $z_{ref}$ is recorded.

*E. Altitude*

Floor level transitions are detected by a separate algorithm that exploits the pressure sensor. Pressure is first converted into altitude and then band-pass filtered to extract altitude variation only. Indeed, it is known [8] that pressure is subject to long time drifts due to meteorological changes, that is why low frequencies components (< 1 minute) have to be removed. When a transition is detected, elevation is compared with floor's heights available within map data, and the floor number is updated.

## III. FLP

*A. Map representation*

Introduction of map is a key aspect of the algorithm design. Hence, map-matching requires estimators supporting hard constraints as well as multi-modal distributions, and the choice of map representation among PM or NM will influence the type of algorithm used to process it.

PM are generally described as grids ([4][5]) or graphs ([8]). Main issues with such a representation are *(i)* the quantization of space which can introduces some errors that accumulates over time, and *(ii)* the map compactness since description occurs at cell's scale, which is typically inferior to 1m (i.e. for instance, a typical office building at CEA has 440 walls on a 6000m² floor, which gives a ratio above 10 between storing cells of 1m² or storing walls). Moreover, when maps are associated with grid based filtering ([4]) or Hidden-Markov Models (HMM, [5]), where the entire map is processed at each time step, computational load could become intractable in real-time, especially if state space includes additional parameters beyond the position itself like (e.g MA), as complexity grows exponentially with the state space dimension. Advantage is that constraints are applied inherently which can save significant processing power with respect to *NM* where constraints checking is the heaviest part of the algorithm.

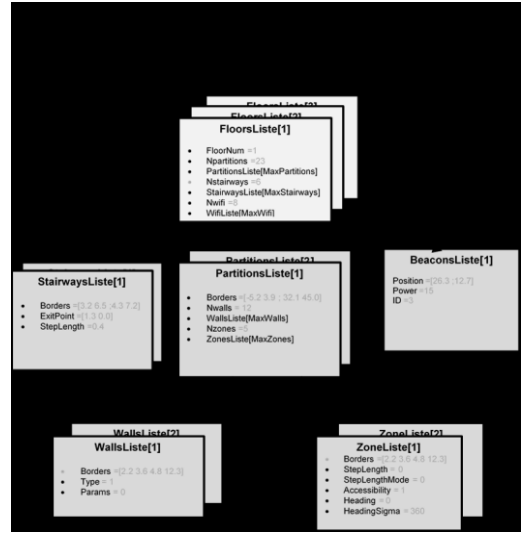

Fig. 2. Overview of the map structure

*NM* typically describes positions of walls represented by segments and each update of the current state (e.g. particle's position) must be checked against walls crossing. This task is computationally intensive because the complexity is of order $N_{particles}.N_{walls}$. Advantage of such a representation is that there is no quantization of space and maps are of small size and easy to generate.

Our approach is based on a *NM* with a particular structure to optimize the constraints checking task and limit the memory usage (see IV.A). Map is organized into 3 levels: map level contains the list of all floors, floor level describes (*i*) positions of the stairways (*ii*) position, ID and power of the Wi-Fi / BLE beacons and (*iii*) a list of partitions which corresponds to small subdivision (typ. 10 to 1000m²) of the floor and aims at reducing both memory usage and overall complexity. Partition level comprises a list of all walls/constraints belonging to the partition, and a list of zones which could be used to describe areas with particular properties (e.g. high accessibility, different step models for moving walkway etc.)

*B. Models*

*1) State model*

Our FLP design is somewhat PDR centric (i.e. sensors are the only modality that is always available), and based on an important observation illustrated on Fig. 3. This example shows a typical PDR trajectory along with ground truth (and estimated FLP trajectory). It is worthwhile noting that the true trajectory could be obtained from PDR trajectory by applying a rotation (i.e. corresponding to MA) and scaling factor. Obviously, some situations could be much more complex (e.g. Fig. 8), nevertheless this assumption is central in our approach. As a consequence, our problem is to estimate the posterior

distribution of the random variable $\mathbf{x}_k=[x_k, y_k, \varepsilon_k, \beta_k]^T$ at time $t_k$ where

- $x_k, y_k$ are the 2D coordinates
- $\varepsilon_k$ is a scaling factor which represents a step length model mismatch and remains almost constant for a given user. Typical values are in range [-0.2 0.2] (i.e. -20% to +20%)
- $\beta_k$ corresponds to MA and is also constant until the device position is changed, or slowly varying because of the gyro drift.

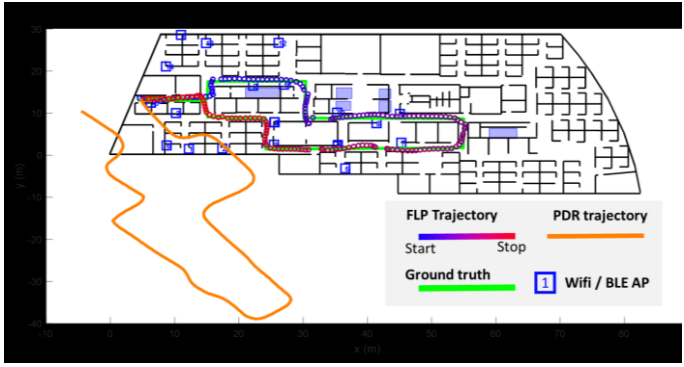

Fig. 3. Example of PDR trajectory

*2) Transition model*

When FLP receives a displacement $L_k$ and a relative heading $\theta_k$ from PDR, state is updated through the transition equation:

$$x_{k+1} = x_k + (1 + \varepsilon_k)(L_k + \eta_k)cos(\alpha_k + \eta_\alpha + \beta_k)$$
$$y_{k+1} = y_k + (1 + \varepsilon_k)(L_k + \eta_k)sin(\alpha_k + \eta_\alpha + \beta_k)$$
$$\varepsilon_{k+1} = \varepsilon_k + \eta_\varepsilon$$ (III-1)
$$\beta_{k+1} = \beta_k + \eta_\beta$$

Where $\eta_\varepsilon$ and $\eta_\beta$, correspond to process noise of the scaling and MA variables, $\eta_d$ and $\eta_\alpha$ correspond to measurement noise of the PDR, all of them being very small (e.g. $\sigma_\varepsilon$= 0.01, $\sigma_\beta$= 0.5deg, $\sigma_d$= 0.1m, and $\sigma_\alpha$= 1deg) according to our previous assumption, independent, centered and normally distributed.

*3) GNSS measurement*

GNSS measurements are described by a very straightforward model assuming Gaussian distribution

$$p(\mathbf{z}_k^{GNSS}|\mathbf{x_k}) = \mathcal{N}(\mathbf{z}_k^{GNSS} - h^{GNSS}(\mathbf{x_k}), \mathbf{\Sigma}^{GNSS}),$$ (III-2)

with

$$h^{GNSS}(\mathbf{x_k}) = [x_k \quad y_k]^T \quad \mathbf{\Sigma}^{GNSS} = \sigma_{GNSS}^2 I_{2\times 2},$$ (III-3)

where $p(\mathbf{z}_k^{GNSS}|\mathbf{x_k})$ refers to the likelihood function of GNSS measurement, $\mathcal{N}(\mathbf{m}, \Sigma)$ normal distribution with mean $\mathbf{m}$ and covariance $\Sigma$ and $I_{2x2}$, identity matrix of size 2x2.

*4) RSS measurement*

Pathloss models, when measurement is expressed in dBm, are generally of the form $z^{RSS} = K - 10\alpha log(d)$, where $d$ stands for the distance between receiver and the transmitting beacon, $\alpha$ is a path loss exponent typically in the range [1 3] for indoors environment (i.e. 2 in free space) and log function comes from the unit (i.e. dBm) in which receivers usually express the RSS value. Since we have a large database at our disposal, comprising up to 257 experiments representing more than 100000 RSS values, we post-processed it to plot the empirical relation between measured RSS versus distance. Objective was to calibrate the log model, but the result (Fig. 4) was very different from our expectations and showed a surprising piecewise linear relation. The flat part of the model, corresponding to distances above 35m can be explained easily by the sensitivity floor-level of the receiver. Indeed, it is likely that RSS measurement bloc of the receiver's front end has limited dynamic (e.g. 40dB) and values below -85dBm are almost saturated. However, regarding the mismatch of the first part of the curve ($d$<35m), we do not have any good explanation so far, our best assumption is that android devices that have been tested (see IV.B) do not return true dBm, but maybe a rescaled version of a linear power.

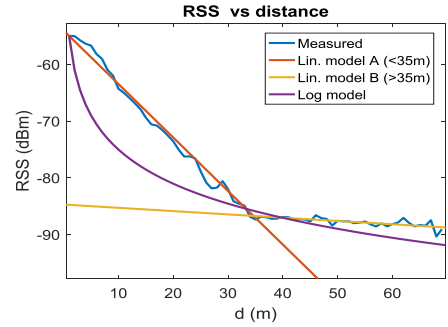

Fig. 4. Experimental curve of rss value vs distance

Our piecewise linear model is

$$h^{RSS}(\mathbf{x_k}, \boldsymbol{b}_i) = \begin{cases} -0.94d(\mathbf{x_k}, \boldsymbol{b}_i) - 54 & d < 35m \\ -0.058d(\mathbf{x_k}, \boldsymbol{b}_i) - 84 & d > 35m \end{cases},$$ (III-4)

where $d(\mathbf{x_k}, \boldsymbol{b}_i)$ is the distance with beacon $\boldsymbol{b}_i$.
We only use scalar measurement for update (only one RSS value per time step) and the likelihood function is given by

$$p(z_k^{RSS}|\mathbf{x_k}) = \mathcal{N}(z_k^{RSS} - h^{RSS}(\mathbf{x_k}), \sigma_{RSS}^2),$$ (III-5)

with $\sigma_{RSS}$ =10dBm.

*C. Particle Filter*

Estimation of the posterior distribution $p(\mathbf{x_k}|\mathbf{z_1}...\mathbf{z_k})$ is performed by a Sequential Importance Resampling (SIR) [17] particle filter (PF), with the difference that only a subset of particles are resampled at each time step. Architecture of the PF is given Fig. 5. Processing is triggered either by a new step ($L$>0) or by a sufficient number of RSS measurements with high values. Then, for each particles, we have the classical prediction step which implements ( III-1), followed by the measurement step (eq. ( III-2) &(III-5) ) and then the collision detection is performed. The prediction step is altered if a DPC has been detected (see II.D). In such a situation, a subset of the particles will resample their MA from a uniform distribution in the range [0 2π], whereas others particles will update their MA value deterministically to preserve the same user heading. In addition, if a particle gets killed during a high-level RSS update, then it gets resampled around the position of the corresponding beacon.

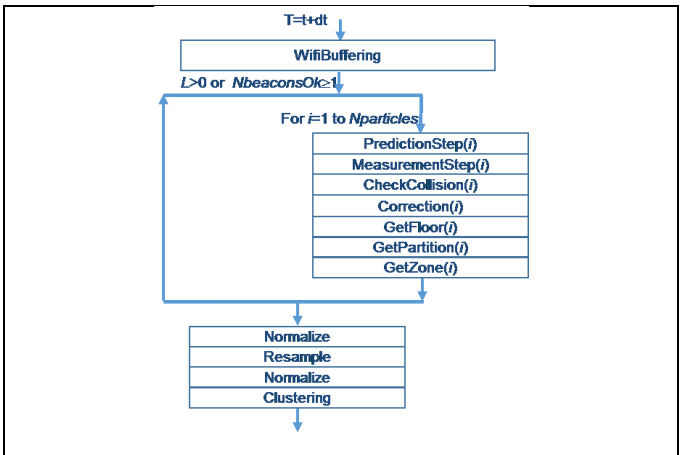

Fig. 5. Overview of the PF's architecture

Collision detection checks if the particle's displacement intersect with walls of the partition the particle belongs to. Objective of map partitioning is to reduce the number of walls that have to be tested. For some collisions (i.e. grazing or head-on collisions) a deterministic correction can be applied to prevent particle from being killed. If a collision is detected and cannot be avoided through correction, then the weight $\omega_k^i$ is set to zero and the particle will be resampled.

When a floor change is detected by a separate algorithm (see §II.E), particles which are not within a stairway zone gets killed, and resampled around a randomly selected stairway exit point, with a probability that depends on the distance between the considered stairway and particle.

Zones are used to indicate high accessibility areas [11], to change the step length model for a particle belonging to a stairway zone or to deactivate GNSS measurements. Indeed, when walking in stairways, length of human steps are equal to the length of stairways' steps and the equation (II-1) is simply replaced by $L = L_{Stairway}$. Also, when user is at the entrance of a building GNSS may be unreliable because of strong multipath or poor satellite visibility. Zones can be used to indicate such bad GNSS conditions.

After each particle has been processed, weights are normalized to sum to unity, and all particles having a weight below a threshold will be resampled.

Because resampling does not preserve the particle's weight, a second normalization stage is necessary. Finally, clustering of the particle's cloud is performed. Indeed, because PF can track multimodal distributions (i.e. several assumptions), averaging over the entire cloud result in solutions that do not make sense A simple example is when the user start in the middle of a straight corridor, because initial heading is unknown, half of the particles will go forth and the other half will go back. Simply averaging over all particles will let the estimated position at the starting point since the center of mass is not moving.

Clustering is based on the famous *k-means* algorithm [18], where the number of clusters (e.g. 5) is set in advance. Normally, *k-means* is an iterative algorithm but here, only one iteration is performed at each time step to reduce computational burden, and we observed in practice that clusters converge to the correct solution since the particle's cloud is stable enough along time. Finally, mean and covariance are computed cluster-wise, and the cluster having the highest weight is outputted as the position estimation.

## IV. EVALUATION

### A. Implementation

Algorithms have been developed under Matlab™ environment and generated in C code using Matlab™ *Embedded Coder*. They have been embedded in an *Android*™ App called *GuideMe* that has been developed for testing, experiment and demonstration purposes. Fig. 6 shows *GuideMe* in *Experiment mode*: the green button on the left is pressed each time the user reaches a landmark defined along a known reference trajectory to construct the ground truth while the device is collecting data. Other items on the bottom serves to define some events, like device position changes. All those annotations can be done remotely by a supervisor through a BLE connection, which is very useful when the device under test is carried in the pocket for instance. At the end of the experiment, a test file is generated automatically that includes both ground truth and collected data (sensors, RSS etc.) to feed a database without any further post-processing. This very simple process allowed us to build a significant database described hereafter.

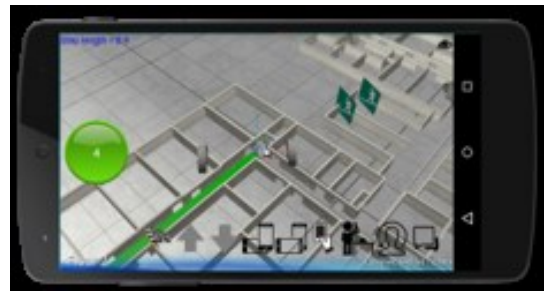

Fig. 6. : Android App called *GuideMe* developped for testing, data acquisition and demonstration purposes

Algorithms have also been integrated into an automated benchmark tool developed in Matlab™ environment that runs the algorithms over the entire database. This tool has also been used for Monte-Carlo optimization of the many parameters to tune, but this is a very heavy process as single trial (over the entire database) takes around 50 minutes on a 12 cores Xeon computer. Nevertheless this optimization has resulted in significant improvement of performances.

Hardware footprint of the FLP has been evaluated on a *BlackFin* 32bits fixed-point platform from Analog Devices™. Although it is not very representative of recent architectures which are much more efficient (e.g. support floating point, multi-core etc.) this choice allows us to have a better control of code execution and accurate cycles count for each part of the algorithm. Results are given in TABLE 1 which indicates that this solution is very light and suitable even for integration in an always-on, low power processor, especially if we consider the large number of particles (1000) and that algorithms are written in non-optimized generated C-code.

TABLE 1. HARDWARE FOOTPRINT OF FLP FOR 1000 PARTICLES

| Code Size | 42.9kB |
|---|---|
| Data Size | 110kB |
| CPU | 45.59MIPS |

Baseline version of the same algorithm had much higher requirements (i.e. 320 MIPS and 861kB of data) but some optimizations helped to reduce them significantly. A first significant gain has been obtained by reducing the update frequency by replacing several small steps by one large steps. A short term, PDR based prediction algorithm, help to compensate for the delay introduced by reducing the update frequency. Then, a more detailed analysis has shown that more demanding parts of the processing were the collision detection stage (180MIPS) and the resampling steps (60MIPS). Resampling step has been optimized by replacing Matlab's random number generator (RNG) based on Merene-Twister algorithm by a linear congruential generator, which is 7.2 times faster. Collision detection computes $N_{particles} \times N_{walls}$ intersections search between two segments (i.e. particle's displacement and wall) involving complex computations (square root and divisions). Basic idea to speed-up the process is to perform a preliminary test that check if (*i*) a wall lies within the cluster bounds and then (*ii*) if horizontal or vertical range of the wall overlaps the horizontal or vertical range of the particle's displacement. Those simple tests, in addition to the partitioning of the map, are based on simple min/max tests and save many more complex intersection search.

Considering data size, main usage is for the map storage. Idea here, is to store locally (i.e. in the SH) only partitions that are not particles empty, whereas the entire map is stored either in the application processor or even remotely. A cache manager system has been implemented to feed the PF with partitions based on particle's positions: when a particle does not belong to any local partitions, a request is send to the application processor (or network) to load new partitions in place of a local partitions that have been marked as empty. If all local partitions contain particles (cache full) then the partition with fewer particles will be replaced and corresponding particles resampled. In practice, cache manager stores only 5 partitions and each partition is sized to a maximum of 100 walls.

*B. Experiments*

Experiments have been conducted over 3 years divided into 8 campaigns of acquisition, resulting 580 files, 240km of walk, about 100 users and 7 different devices. However first campaigns were focusing on PDR envelopments and are not suitable for FLP testing, because they are mainly outdoor with very simple trajectories.

Considering only subset of the database that is relevant for FLP testing we still have 257 files corresponding to 4 campaigns, 117km and 60 users described in Table 2. Those experiments took place in 4 different sites: 3 office type in France and in the US with many rooms and narrow corridors, 1 mall type (mix of corridors and large open spaces). One site (CEA Grenoble, France) comprises 4 floors connected with both stairways and elevators. We used Nexus 4, Nexus 5, Galaxy Note 3 and one proprietary platform equipped with consumer grade sensors.

One typical experiment last about 10 minutes or more and the user is asked to follow a predefined path and, for some experiments, he is also asked to do some additional stuff like making a phone call, changing the smartphone position, or find a specific information on a poster (i.e. to have some trampling behavior). All the classical device positions have been tested (portrait, landscape, call, pocket, chest, waist, backpack and swinging in hand). Typical density of radio beacons (when present) is 1 per 300m², placed at positions that are independent of algorithm's performances.

User is followed by a supervisor that indicates directions and also annotates when user passes a reference landmark which are typically located at each turn. Fig. 7 shows a typical example of experiment at CEA (campaign #7), with several floors, various sensors availability (zones with GNSS, zones with Wi-Fi, stairways and elevator), a duration of about 13 minutes and a length of 575m. FLP's trajectory starts in blue and ends in red, Ground truth is displayed in green.

*C. Metric*

The metric that has been used for performance evaluation is the percent of time when error is inferior to 5m or 10m (D5 or D10 resp.) in addition to root-mean-square error (RMSE). Reason for this choice is related to the behavior of PF : when a wrong solution is chosen by the algorithm (e.g. wrong corridor) it can result in very large errors that could grow RMSE value significantly whereas user experience would be mostly affected by how often the FLP get lost.

In practice, it is seldom that algorithms achieve D5=100%, even when the result looks perfect, for several reasons : linear interpolation between two reference landmarks is not very accurate, user can walk far from the predefined path especially outdoor, ground truth is not well defined in stairways or because position output is slightly delayed wrt ground truth. We arbitrarily choose to classify experiments according to their D5 value as : *perfect* ( 80% < D5), *Good* (60%< D5 < 80%) , *Middle* (40% < D5 < 60%) and *Bad* (D5 < 40%).

*D. Results*

Results are summarized in Table 2. Worst results are obtained for campaign #6 (denoted as C6) where no Wi-Fi access point (AP) have been used. Hence, entire trajectory estimation relies on PDR and map only which is a very tough use case considering the duration (about 17mn) and the number of floors. If, at a certain point, FLP get lost, it has no chance to recover and will remain in this state for the rest of the trajectory. C7/UC1 is almost the same as C6 except that six Wi-Fi APs have been added on a small part of the trajectory (see Fig. 7). This improves significantly the D5 performance from 23% without AP to 50% when using a few beacons, because FLP can be reset to a good position when he enters the Wi-Fi zone. It should be noted that performances for this campaign are a little bit better than it seems, because this campaign concentrates all metric issues mentioned previously, especially the outdoor section and numerous floor changes. For instance, D5 performance of example given in Fig. 7 is 63% only, whereas estimated trajectory looks close to the ground truth almost everywhere. Only outdoor zone is bad, but this is mainly due to GNSS performance at the foot of a five floors building. C8/UC1 is somewhat similar to C7/UC1 but with different trajectory, no outdoor zone, more APs and with device position change during the trajectory. Increasing the number of beacons improves D5 from 58% to 73% despite a more difficult scenario due to the DPC, and almost all experiments are *good* (56.7%) or *perfect* (30%).

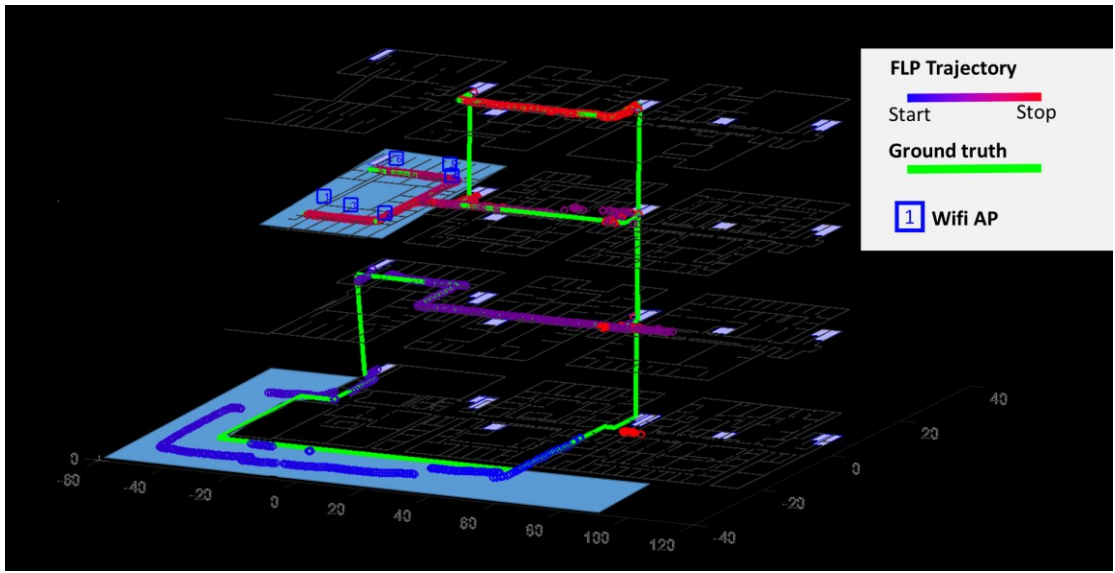

Fig. 7. Example of typical trajectory with sevreal floors, outdoor and Wi-Fi zone

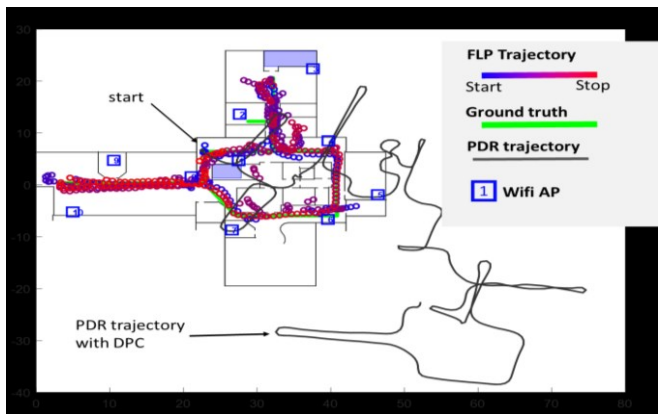

Fig. 8. Example of FLP trajectory from C8/UC2, with single floor, good WI-Fi / BLE coverage and DPC

When trajectories are single floor, with good Wi-Fi/BLE coverage in office type environments (corresponding to C8/UC2, C9/all use cases) then performances are very good with around 73% of experiments falling into the *perfect* category with an average RMS error of 4.3m even with DPC. Fig. 8 gives an example of such a trajectory and also the associated (input) PDR trajectory which is strongly affected by DPC. However, FLP is only slightly disturbed by DPC, and manages to find quickly the correct path thanks to map and RSS information.

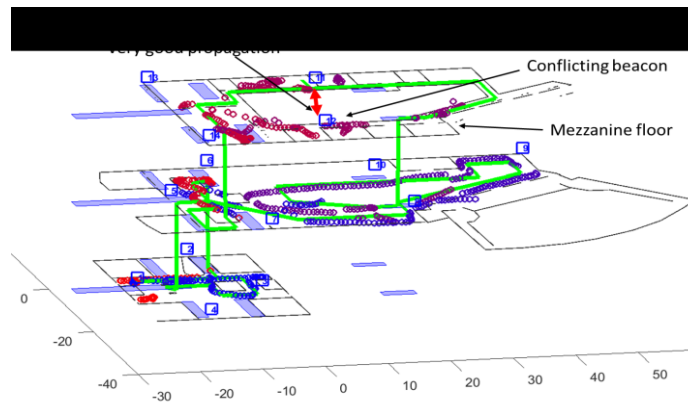

Fig. 9. Exemple of difficult use case (C8 / UC3) with a difficult map topology having large open spaces and long stairways

Assuming a standard Wi-Fi/BLE coverage (1 beacon per 300m²), limitations of our approach are observed for maps having a complex topology as shown on Fig. 9. This venue has large open spaces on first floor and long stairways which are poorly handled by our algorithms. In addition, the 2$^{nd}$ floor is a mezzanine and it has been observed that some beacons can be received at high levels at the opposite side of the building thanks to a very good, free space propagation. For this kind of complex scenario (C8 / UC3) D5 performance drop to 48.7%, and majority of experiments (70%) are classified as *middle*.

Finally, the overall performance computed on the entire database gives D5 of 70% with large disparities depending on the use case that is considered. FLP is not working for about 10% of the experiments, mainly represented by C6 where positioning relies on map and sensors only, and where no absolute reset is possible (e.g. using RSS or GNSS). Systems is working very well for 40% of the use cases, which typically correspond to single floor and standard Wi-Fi / BLE coverage (1 beacon per 300m²), even with DPC while the user is walking. Performances start to decrease when the path goes through different floors (23% of good) because the stairways transitions are not well modeled and pressure sensor introduces some delays. If, in addition, the venue has a complex map with large open spaces for instance, then performances are middle (around 25%).

TABLE 2. DATABASE SUMMARY AND RESULTS

| Camp. | Use Case | Place | # files | Dur. (mn) | Dist. m | Outdoor | # floors | Wifi / BLE | DPC | D5 (%) | D10 (%) | RMS (m) | Bad (%) | Middle (%) | Good (%) | Perfect (%) |
|---|---|---|---|---|---|---|---|---|---|---|---|---|---|---|---|---|
| #6 | UC1 | CEA | 10 | 17 | 972 | no | 4 | 0 | no | 23,76 | 33,24 | 39,96 | 90,0 | 10,0 | 0,0 | 0,0 |
| #7 | UC1 | CEA | 40 | 13 | 575 | yes | 4 | 6 | no | 50,68 | 71,24 | 13,74 | 15,0 | 72,5 | 12,5 | 0,0 |
|  | UC2a | CEA | 20 | 10 | 841 | no | 1 | 6 | no | 83,75 | 96,65 | 7,60 | 0,0 | 0,0 | 25,0 | 75,0 |
|  | UC2b | CEA | 20 | 10 | 841 | no | 1 | 6 | yes | 77,04 | 89,91 | 10,82 | 5,0 | 5,0 | 55,0 | 35,0 |
| #8 | UC1 | CEA | 30 | 13 | 476 | no | 4 | 20 | yes | 72,89 | 86,89 | 7,40 | 0,0 | 13,3 | 56,7 | 30,0 |
|  | UC2 | IFR | 30 | 10 | 384 | no | 1 | 10 | yes | 82,04 | 95,20 | 4,68 | 0,0 | 0,0 | 26,7 | 73,3 |
|  | UC3 | WTC | 30 | 10 | 347 | no | 3 | 14 | yes | 48,75 | 75,76 | 10,56 | 16,7 | 70,0 | 13,3 | 0,0 |
| #9 | UC1 | CEA | 35 | 13 | 254 | no | 1 | 35 | Yes | 84,42 | 91,57 | 5,19 | 5,7 | 5,7 | 17,1 | 71,4 |
|  | UC2 | IFR | 25 | 3 | 130 | no | 1 | 21 | Yes | 87,20 | 96,85 | 3,29 | 0,0 | 16,0 | 8,0 | 76,0 |
|  | UC3 | ISJ | 17 | 3 | 147 | no | 1 | 30 | yes | 81,29 | 90,79 | 4,20 | 5,9 | 11,8 | 11,8 | 70,6 |
| Total |  |  | 257 | 44h20 | 117km |  |  |  |  | 70,46 | 84,90 | 9,07 | 9,3 | 24,9 | 23,3 | 42,4 |

## V. CONCLUSION

A positioning algorithm that makes use of various data available in a smartphone has been presented. This algorithm addresses some of the most challenging use cases (e.g. DPC) and is intended for a lightweight, always-on, hardware integration. Implementation on a 32bits, fixed-point platform (i.e. *BlackFin* processor) has shown a hardware footprint requiring processing power of 45MIPS, and memory size of 150kB (code+data), which is suitable for SH integration. Extensive testing over more than 250 experiments representing various conditions, gives an overall accuracy of 9m RMSE, which improves to 4.3m for favorable scenarios. Analysis of those two figures reveals that algorithm is working very well for some environments (e.g. narrow corridors, good Wi-Fi / BLE coverage) even with DPC, but deteriorates under one or several of these factors : (*i*) without any RSS measurements, (*ii*) in large open spaces, (*iii*) in stairways transitions.